\documentclass{article}

\def\*{{\bf ***}}

\def\a{\alpha}
\def\b{\beta}

\def\ga{\gamma}
\def\de{\delta}   

\def\phi{\varphi}

\def\s{\sigma}
\def\z{\zeta}
\def\om{\omega}
\def\th{\theta}
\def\vth{\vartheta}

\def\C{{\bf C}}

\def\h{{\cal H}}

\def\L{{\cal L}}

\def\N{{\cal N}}
\def\Mb{{\bf M}}

\def\R{{\bf R}}

\def\T{{\rm T}}
\def\V{{\cal V}}

\def\Ga{\Gamma}

\def\La{\Lambda}
\def\Om{\Omega}

\def\pa{\partial}

\def\d{{\rm d}}       
\def\w{\wedge}
\def\xb{{\bf x}}
\def\yb{{\bf y}}

\def\o+{\oplus}

\def\ss{\subset}

\def\<{\langle}
\def\>{\rangle}

\def\interno{\hskip 2pt \vbox{\hbox{\vbox to .18
truecm{\vfill\hbox to .25 truecm
{\hfill\hfill}\vfill}\vrule}\hrule}\hskip 2 pt}

\def\({\left(}
\def\){\right)}
\def\[{\left[}
\def\]{\right]}
\def\=#1{\bar #1}
\def\~#1{\widetilde #1}
\def\.#1{\dot #1}
\def\^#1{\widehat #1}
\def\"#1{\ddot #1}

\def\ref#1{\cite{#1}}

\def\Remark#1{\medskip \noindent {\bf Remark {#1}}}

\begin{document}

\title{\bf A variational principle for \\ volume-preserving dynamics}

\author{Giuseppe Gaeta\footnote{Supported by ``Fondazione CARIPLO per la ricerca scientifica''}\footnote{g.gaeta@tiscali.it}\\ 
{\it Dipartimento di Matematica, Universit\'a di Milano,} \\ 
{\it via Saldini 50, I--20133 Milano (Italy)} \\
{~~} \\
Paola Morando\footnote{morando@polito.it} \\ 
{\it Dipartimento di Matematica, Politecnico di Torino,} \\
{\it Corso Duca degli Abruzzi 24, I--10129 Torino (Italy)} }

\date{~}

\maketitle

\noindent{\bf Summary.} We provide a variational description of any Liouville (i.e. volume preserving) autonomous vector fields on a smooth manifold. This is obtained via a ``maximal degree'' variational principle; critical sections for this are integral manifolds for the Liouville vector field. We work in coordinates and provide explicit formulae.

\section*{Introduction}

It is well known that Hamiltonian dynamics preserves the volume in phase space (Liouville theorem); they are thus a prominent example of {\it incompressible} (or volume-preserving) dynamics. 

However, other dynamics which preserve the volume, or more generally a measure, in phase space without being necessarily Hamiltonian, are also of physical interest. These are called {\it Liouville vector fields} (see below for a precise definition). The motion of an incompressible fluid is of course described by volume-preserving dynamics, and is a prominent physical example of Liouville dynamics.

Among relevant classes of volume-preserving dynamical systems which are not necessarily hamiltonian, we mention Nambu mechanics \cite{Nam} and reversible dynamics \cite{LaR}. 

It has long been known that several subclasses of Liouville dynamics share important features with Hamiltonian one; for example, statistical mechanics can be properly based on Nambu dynamics (this was actually the motivation for it \cite{Nam}), and reversible vector fields share many results of the perturbation theory of Hamiltonian systems, including KAM theory \cite{LaR}.

It is thus entirely natural, and justified by physical relevance, to wonder what is the extent to which Liouville dynamics shares the structures and properties of Hamiltonian one.

The question has been studied by several groups; we mention here in particular the work of \cite{Mar}, extending to Liouville dynamics several geometric structures -- including the Poisson bracket formulation -- of Hamiltonian dynamics, and clarifying several points in the geometrical formulation of Liouville dynamics; see \cite{Ma2,Ma4} for further developements in this direction (also connected with Nambu-type systems). 

Hamiltonian dynamics can be characterized in terms of a variational principle, and this is instrumental to many results in hamiltonian theory; it is thus natural to investigate if some kind of variational formulation is also possible for general Liouville vector fields (in general, a standard variational formulation is not possible).

The purpose of this note is to answer this question in the positive: we find that any Liouville vector field on the phase space manifold $P$ can be described as the unique properly normalized characteristic vector field of a maximal degree variational principle (these notions will be defined below) on the extended phase space $M = \R \times P$; this applies in particular to autonomous Liouville vector field on $P$, in which case the $\R$ factor can be thought as the physical time.

It should be stressed that, although in this note we work in local coordinates, an intrinsic discussion would be possible and is actually given elsewhere, leading to more general results \cite{GMvar}. Such an approach, however, requires to use the Cartan theory of exterior differential systems; this theory is well known in differential geometry and in the geometric theory of differential equations, but we believe the results given here can be of interest to all physicists working on incompressible dynamics, which in general are not necessarily familiar with Cartan theory. 

Luckily, the case of Liouville fields can be dealt with at the much  simpler level considered here, i.e. working in coordinates and involving only basic notions of differential geometry and rather elementary mathematical analysis. 

\bigskip\noindent
{\bf Acknowledgement.} The work of GG has been supported by {\it ``Fon\-da\-zio\-ne CARIPLO per la ricerca scientifica''} under the project {\it ``Teoria delle perturbazioni per sistemi con simmetria''}.

\section{Liouville dynamics}

\subsection{Liouville vector fields on phase space}

Let $P$ be a smooth and orientable $N$-dimensional manifold, from now on called the ``phase space''; we will, as customary, denote by $\La (P)$ the set of differential forms on $P$, and $\La^k (P) \ss \La (P)$ denotes the set of forms of degree $k$. 
Recall that all volume forms on $P$ (also referred to as measures) are equivalent; we choose a reference volume form $\Om$ on $P$. 

Let $X$ be a vector field on $P$; we say that $X$ is a Liouville vector field (with respect to the measure $\Om$) if it preserves $\Om$, i.e. if $\L_X (\Om ) = 0$, with $\L$ the Lie derivative.

As $\L_X (\a) := (X \interno \d \a) + \d (X \interno \a)$ and $\d \Om = 0$, $X$ is Liouville w.r.t. $\Om$ if and only if $X \interno \Om$ is
closed; locally this means that there exists a (N-2)-form $\ga$ such
that
$$ X \interno \Om \ = \ \d \ga \ . \eqno(1) $$
Thus $X$ is {\it Liouville with respect to $\Om$} iff
for any neighbourhood $A \subseteq P$, eq.(1) is verified for some $\ga \in \Lambda^{N-2} (A)$, i.e. iff $X \interno \Om$ is exact on $A$.

We say that $X$ is {\it globally Liouville with respect to $\Om$} iff (1) is verified for some $\ga \in \Lambda^{N-2} (P)$, i.e. iff $X
\interno \Om$ is exact on $P$.

In the following, for ease of discussion and notation, we will deal with the case where $X$ is globally Liouville; however it is clear that for the sake of local considerations the two cases are equivalent.

Note that if $X$ and $\Om$ are given, $\ga$ is not uniquely defined by (1) (we can always add a closed form $\ga_1$); on the other hand, if $\Om$ and $\ga$ are given, then (1) uniquely defines $X$.

We have chosen a given volume form $\Om$; for any different choice $\~\Om$ of the volume form we have $\Om = \rho \~\Om$, with $\rho$ a nowhere vanishing (positive, to preserve orientation) function $\rho \in \La^0 (P)$; moreover if $Y \interno \~\Om = \d \ga$ and $\Om = \rho \~\Om$, then $X := \rho Y$ satisfies (1) with the same $\ga$. Hence whenever we have a vector field $Y$ which is Liouville with respect to a measure $\~\Om$, we can reduce to considering $X$ and an exact volume form. Thus, from now on we will assume $\Om = \d \s$, for ease of computation.

\subsection{Enlarged phase space}

It is convenient to consider the ``enlarged phase space'' $ M := \R \times P$, where the $\R$ factor corresponds to the time coordinate $t$; we denote the reference volume form in $M$ corresponding to $\Om \in \La^N (P)$ by $\Om_M := \d t \w \Om \in \La^{N+1} (M)$. Then the dynamics defined by $X$ in $P$ corresponds to the flow of $Z$ in $M$, where
$$ Z \ := \ \pa_t \, + \, X \ ; \eqno(2) $$
this satisfies, by construction, $Z \interno \d t = 1$.

We consider now a form $\vth \in \Lambda^{N-1} (M)$ built from the $\s$ defined above and from the form $\ga$ associated to $X$ via (1); this is defined as 
$$ \vth \ := \ \s \ + \ \d t \w \ga \ ; \eqno(3) $$
it is immediate to check that $\d \vth $ is nowhere zero. 

We now prove that the form $\vth$ defined in (3) defines a {\it  unique} vector field $\~Z$ on $M$ via 
$$ \~Z \interno \d \vth \ = \ 0 \ \ ; \ \ \~Z \interno \d t \ = \ 1 \ \ . \eqno(4) $$
Note that $\~Z \interno \d t \not= 0$ implies that the vector field $\~Z$ has a nonzero component along $\pa_t$. 
\medskip

To see that (4) defines a unique field, it suffices to note that $\d \vth$ is a $N$-form in the manifold $M$ of dimension $N+1$: in this case its annihilator $\N (\d \vth)$ (the set of vector fields $Y$ on $M$ such that $Y \interno \d \vth = 0$) is a one dimensional module over $\Lambda^0 (M)$. Obviously the first of (4) amounts to the requirement $\~Z \in \N (\d \vth)$. 

\Remark{1.} More precisely, any form $\a \in \Lambda^{N} (M)$ can be written in local coordinates $(x^0, x^1,...., x^{N})$ (we use the notation $\pa_\mu \equiv \pa / \pa x^\mu$ and $t = x^0$) as $A^\mu (x) (\pa_\mu \interno \Om_M)$; it is easy to check that the vector fields in $\N (\a)$ are then written as $Y = f^\mu \pa_\mu$ with $f^\mu (x) = F(x) A^\mu (x)$, where $F \in \Lambda^0 (M)$. $\odot$
\medskip

Note next that $Y \in \N (\d \vth)$ (and $Y \not= 0$) is necessarily such that $Y \interno \d t \not= 0$, just by the form of $\vth$. The second equation in (4) is therefore just a normalization condition, selecting a unique vector field $\~Z$ out of the one dimensional module $\N (\d \vth)$.

Actually, the vector field $\~Z$ defined by (4) is just the vector field $Z$ defined above in (2). Given that $\~Z$ is unique, to prove that $\~Z \equiv Z$ it suffices to check that $Z$ satisfies (4).

It is obvious that $Z \interno \d t = 1$, as already remarked. As for the first of (4), we have, using $\d \s = \Om$ and (1),  
$$ \begin{array}{rl}
Z \interno \d \th \ =& \ X \interno \Om \ + \ (-1)^N \, (Z  \interno \d \ga) \w \d t \ + \ (-1)^{(2N-1)} \, \d \ga \\
 =& \ + \, \d \ga \ + \ (-1)^N \, (X \interno \d \ga) \w \d t \ - \ \d \ga \\
 =& \ (-1)^N \, (X \interno \d \ga ) \w \d t \ = \ 
(-1)^N \, [ X \interno (X \interno \Om)] \ = \ 0  \ . \end{array} $$
This completes the proof; we summarize our results as follows.

\medskip\noindent
{\bf Lemma 1.} {\it The equations (4), with $\vth$ given by (3), select uniquely the vector field $Z$ given in (2), where $X$ satisfies (1).}
\medskip

\section{Variational setting}

We want now to prove that $Z$, and hence $X$, can be given a variational characterization. This will be based on a suitable  fibration $\~\pi : P \to Q$ of the phase space over a smooth $K$-dimensional manifold $Q$, which induces a fibration $\pi : M \to B$ of the extended phase space, $\pi := id \times \~\pi$. In order to do this we will have to introduce the concept of {\it maximal degree} variational principle, and of the associated {\it characteristic field}.

\subsection{Variational principles}

Consider a general smooth bundle $\pi : E \to B$, with $E$ a smooth manifold of dimension $n$ and $B$ a smooth manifold of dimension $k$, with $1 \le k < n$ (in the next section we will specialize to the case $E = M = \R \times P$). We assume the fibers $\pi^{-1} (x)$ are parallelizable.

We denote by $\Ga (\pi)$ the space of smooth sections $\phi : B \to E$ of this bundle, and by $\V (\pi)$ the space of vector fields on $E$ which are vertical for $\pi$, i.e. tangent to fibers $\pi^{-1} (x)$. Given a form $\a \in \Lambda (E)$, we denote its pullback by a section $\phi \in \Ga (\pi)$ as $\phi^* (\a)$.

Let $D \ss B$ be a domain (i.e. a closed compact manifold with boundary) in $B$, and $\eta \in \Lambda^k (E)$; we define a functional $I : \Ga (\pi ) \to \R$ by
$$ I (\phi) \ := \ \int_D \phi^* (\eta ) \ . \eqno(5) $$

A vector field $V \in \V (\pi)$ obviously induces an action on $\Ga (\pi)$; this results in turn into an action on $I (\phi)$. More precisely, denote by $\psi_s$ the local flow of $V$ on $E$, and consider a section $\phi_0 \in \Ga (\pi)$; the flow in $\Ga (\pi)$ originating from $\phi_0$ is the one-parameter family of local sections  $\phi_s = \widetilde{\psi}_s (\phi_0):= \psi_s \circ\phi_0$. 
We define the variation of $I$ under $V$ as
$$ (\de_V I_D ) (\phi) \ := \ {\d ~ \over \d s} \ \[ \int_D \, \( \widetilde{\psi}_s (\phi) \)^* (\eta) \]_{s=0} \ . \eqno(6) $$ 

We denote by $\V_D (\pi) \ss \V (\pi)$ the set of vector fields vertical for $\pi$ which vanish on the fibers over $\pa D$. 
We say that $\phi \in \Ga (\pi)$ is {\it extremal} (or also {\it critical}) for $I$ defined by (5) if and only if $\de_V I (\phi) = 0$ for all $V \in \V_D (\pi)$.

It is well known that, equivalently, $\phi$ is critical for $I$ defined by (5) if and only if 
$$ \phi^* (V \interno \d \eta ) \ = \ 0 \ \ \ \ \ \forall V \in \V_D (\pi) \ . \eqno(7) $$
Given the complete equivalence of this condition with the previous definition \cite{Her,Sau}, we can take (7) as the definition of critical section.

It will be convenient, as our considerations will be local, to use local coordinates in $E$ adapted to the fibration $\pi : E \to B$; that is, we use coordinates $(x^1 , ... , x^k ; y^1 , ... , y^p )$ where $k+p=n$, $(x^1,...,x^k)$ will be coordinates in $B$, and $(y^1,...,y^p)$ are vertical coordinates spanning the fibers $\pi^{-1} (x)$. We write, for ease of notation,
$$ \pa_i \ := \pa / \pa x^i \ \ , \ \ \pa_\a \ := \ \pa / \pa y^\a \ \ ; \ \ i=1,...,k \ , \ \a = 1,...,p \ . \eqno(8) $$
The image of a section $\phi \in \Ga (\pi)$ is then locally the graph of a function $u : \R^k \to \R^p$, i.e. $\phi = \{ \xb , \yb = u (\xb ) \}$. 
In this coordinates, $ \phi^* (\d y^\a ) = (\pa u^\a / \pa x^i ) \d x^i $.

The determination of equations describing critical sections for $I$ is very well known, and we just recall it to fix notation. Any $V \in \V (\pi)$ is written in these local coordinates as $V = f^\a \pa_\a$; the set $\V_D (\pi)$ is identified by the condition that $f^\a = 0$ on $\pi^{-1} (\pa D)$. Thus (7) requires $\phi^*[f^\a (\pa_\a \interno \d \vth )] = 0 $ for arbitrary $f^\a$ (vanishing on $\pi^{-1} (\pa D)$); hence we need that $ \phi^* (\pa_\a \interno \d \vth) = 0$ for all $\a =1,..,p$. 

It is natural to introduce the $k$-forms 
$ \Psi_\a := \pa_\a \interno \d \vth$; the section $\phi$ is critical for $I$ if and only if 
$$ \phi^* (\Psi_\a ) \ = \ 0 \ \ \  \a=1,...,p \ . \eqno(9) $$ 

\Remark{2.} 
It should be noted that the fibration $\pi : E \to B$ is nontrivial for $1 \le k \le n-1$; however, the case $k = n-1$ will provide trivial equations (as discussed in \cite{GMvar}); thus we have to require $1 \le k \le n-2$. 
The case $k=1$ corresponds to the well known case of one independent variable (as in standard Hamilton dynamics), and it is well known that in this case the variational principle identifies a vector field on $E$, i.e. yields a system of ODEs for $\phi$. 
For $k>1$, the variational principle yields a system of PDEs for $\phi$; in local coordinates as those introduced above, this is a system of $p$ equations for the $p$ functions $u^\a (x^1,...,x^k)$ depending on $k$ independent variables.
$\odot$ 

\Remark{3.}
Note that the fibration $\pi: E \to B$ is to a large extent arbitrary; basically it amounts to a choice of which variables should be considered as independent ones and which as dependent in formulating the variational principle, and has no intrinsic meaning. Later on, when we specialize to $E = M = \R \times P$ and we study a dynamics (see sect.3), we will require that the time is kept as one of the independent variables: this means that $B = \R \times Q$, and the fibering $\pi : M \to B$ is induced by a fibering $\~\pi : P \to Q$ as anticipated above. Note also that the variational principle is always formulated over a domain $D \ss B$, i.e. in local terms. 
$\odot$

\subsection{The maximal degree case}

In view of remark 2 above, we say that for $k = n-2$ we have a {\it maximal degree variational principle}. 
We are interested in this case: we will show that these variational principles still identify a vector field on $E$, i.e. a system of ODEs associated to it.

As $k=n-2$, we have $p=2$; we will write $z \equiv y^1$, $w \equiv y^2$ to avoid a plethora of indices.
We write $\om = \d x^1 \w ... \w \d x^k$ for the reference volume form in $B$; the reference volume form in $E$ will of course be $\pi^* (\om) \wedge dz \wedge dw$; in the following we will write, with a slight abuse of notation, $\om$ for $\pi^* (\om)$.
In this case, when studying the variational principle defined by (5), (6), we should consider $\beta := \d \eta \in \Lambda^{n-1} (E)$; we can always write any $\beta \in \Lambda^{n-1} (E)$ in the form
$$ \begin{array}{rl}
\beta \ =& \ \sum_{\mu=1}^k A^\mu \[ \om_{(\mu)} \w \d z \w \d w \] \ + \\ & \ + \ (-1)^k f \[ \om \w \d w \] \ + \ (-1)^{k+1} g \[ \om \w \d z \] \ , \end{array} \eqno(10) $$
with $\mu=1,2$, $A^\mu,f,g$ smooth functions of $({\bf x},z,w)$, and  $\om_{(\mu)} := \pa_\mu \interno \om$.

In the following, we will assume that the vector ${\bf A} = (A^1 , ... , A^k)$ is not identically zero; in this case we say that the maximal degree variational principle defined by $\eta$ is {\it proper}.

In the present notation, we choose $\pa_z$ and $\pa_w$ as generators of $\V (\pi)$, i.e. $\Psi_1 = \pa_z \interno \beta$, $\Psi_2 = \pa_w \interno \beta$. With $\phi \in \Ga (\pi)$, we have 
$$ \begin{array}{rl}
\Psi_1 \ =& \ (-1)^{k-1} \, \[ A^\mu \, (\om_{(\mu)} \w \d w ) \ + \ (-1)^k g \,  \om \] \ ; \\ 
\Psi_2 \ =& \ (- 1)^k \, \[ A^\mu \, ( \om_{(\mu)} \w \d z) \ + \ (-1)^k f \, \om \] \ . \\ 
\phi^* (\Psi_1) \ =& \  \phi^* \[ A^\mu (\pa w / \pa x^\mu) \, - \,  g \] \, \om \ ; \\ 
\phi^* (\Psi_2) \ =& \ - \, \phi^* \[ A^\mu (\pa z / \pa x^\mu) \, - \, f \] \, \om \ . \end{array} \eqno(11) $$

Requiring the vanishing of both $\phi^* (\Psi_j)$ for $j=1,2$ means looking for solutions of two quasilinear first order PDEs, i.e. 
$$ \phi^* \[ \L_Y (z) \, - \, f \] \ = \ 0  \ \ \ ; \ \ \ \phi^* \[ \L_Y (w) \, - \, g \] \ = \ 0  \eqno(12) $$
(with obviously $Y = A^\mu \pa_\mu$, and $\L_Y$ the Lie derivative). Note that as ${\bf A} \not= 0$ in (10), we are guaranteed $Y \not= 0$.

The relevant property is that the equations can be written in terms of the action of the same (nonzero) vector field $Y$, or more precisely \cite{Arn} in terms of the (non vertical, as $Y \not= 0$) vector field $W = Y + f \pa_z + g \pa_w$ on $E$, i.e.  
$$ W \ = \ \sum_{\mu=1}^{n-2} \, A^\mu (\xb;z,w) \, {\pa \over \pa x^\mu } \ + \ f (\xb;z,w) \, {\pa \over \pa z} \ + \ g (\xb;z,w) \, {\pa \over \pa w} \ . \eqno(13) $$

Indeed, as well known (see e.g. chapter II.7.G of \cite{Arn}), the $\R^2$-valued function $u(x,t) = \left( z(x,t) , w (x,t) \right)$ is a solution to the system of quasilinear PDEs (12) if and only if its graph is an integral manifold for the associated characteristic system
$$  d x^\mu / d s = A^\mu \ , \ d z / d s =  f \ , \ d w / d s  = g \ , \eqno(14) $$ 
i.e. for the $W$ given above (the characteristic system (14) is often written in the so-called symmetric form, i.e. as $d x^1 / A^1 = ... = d x^k / A^k = d z / f = d w / g$).

We recall that if $X$ is a vector field on $E$ and the submanifold $S \ss E$ is such that $X (x) \in \T_x S$ for all $x \in S$, we say that $S$ is an {\it integral} (or {\it invariant}) {\it manifold} for $X$; if $S$ is invariant for $X$ and one-dimensional, we also say it is an {\it integral curve} of $X$.

It is thus entirely natural to call the $W$ given by (13) the {\it  characteristic vector field} for the maximal degree variational principle on $\pi : E \to B$ defined by a form $\eta \in \La^{n-2} (E)$ such that $\d \eta = \b$ is given by (10).

Note that the above computations imply that for given $\eta$ we have a one dimensional module of characteristic vector fields (all differing by multiplication by a nowhere zero smooth function), defining a unique direction field on $E$ (see \cite{Arn}). 

Summarizing, with the above discussion we have proved that:

\medskip\noindent
{\bf Theorem 1.} {\it Let $\pi : E \to B$ be a fibration of the smooth $n$-dimensional manifold $E$ over the $(n-2)$-dimensional manifold $B$, and let $\eta \in \Lambda^{n-2} (E)$ be such that $\d \eta = \beta$ is written as in (10). Then the section $\phi \in \Ga (\pi)$ is critical for the maximal degree proper variational principle defined via the functional (5) if and only if it is an invariant manifold of the characteristic vector fields for the variational principle.}

\medskip\noindent
{\bf Corollary.} {\it Given an integral curve $\Ga$ of a characteristic vector field for the variational principle defined by (5) and a critical section $\phi$ of it, either $\Ga \cap \phi = \emptyset$ or $\Ga \ss \phi$.}\medskip

The corollary implies that we can describe the $k$-dimensional critical sections $\phi$ for a variational principle as the union of integral curves of the characteristic vector field $W$ for the same variational principle passing through a suitable submanifold $\phi_0 \ss \phi$ of dimension $k-1$.

Note that if one of the $A^\mu$, say $A^1$, is nowhere zero, we can divide this out from $W$, and obtain a vector field of the form 
$$ Z \ = \ \pa_1 \ + \ X \eqno(15) $$
with $X \interno \d x^1 = 0$. 

We would also like to remark that $W$ is tangent to sections $\phi \in \Ga (\pi)$ such that $\phi^* (\Psi_1) = 0 = \phi^* (\Psi_2)$, see above and \cite{Arn}; on the other hand, $\phi^* (\Psi_i) = 0$ means that $\Psi_i$ vanish on vector fields tangent to $\phi$, hence vanish if evaluated on $W$, i.e. $W \interno \Psi_i = 0$. We have thus proven that:

\medskip\noindent
{\bf Lemma 2.} {\it The $W$ identified by (13) satisfies $W \interno \Psi_1 = 0 = W \interno \Psi_2$.}

\section{Variational principle for Liouville dynamics}

In this section we apply the discussion on maximal degree variational principles developed in the previous section, to the case of Liouville dynamics.

We consider, with the notation introduced in section 1, $E = M = \R \times P$, and $B = \R \times Q$; note that $n = N+1$ and $k = K+1$. We stress that the fibering $\pi : M \to B$ is naturally induced by a fibering $\~\pi : P \to Q$, i.e. with obvious notation $\pi = id \times \~\pi$ (see also remark 3 above).

Accordingly, we will use local coordinates adapted to the double fibration of $M$; we call these $(t; \xi^1,...,\xi^k ; y^1 , y^2)$, where $t$ is the coordinate along $\R$, $(\xi^1,...\xi^k)$ are coordinates in $Q$ -- hence $(t;\xi)$ are coordinates in $M$ -- and $y^\a$ are ``vertical'' coordinates on the fibers $\pi^{-1} (t;\xi) = \~\pi^{-1} (\xi)$. The setting of the previous section is recovered via $x^1 := t $, $x^{j+1} := \xi^j$. 

In order to recover the Liouville vector field $X$ satisfying (1) as the vector field identified by a maximal degree variational principle, it suffices to choose as $\eta$ the form $\vth$ defined in (3). Note that in this case, using the notation (10), we have indeed that $A^1$ is nowhere zero.

With local coordinates $(\xb,z,w)$ as in the previous section, write $V_1 = \pa_z$, $V_2 = \pa_w$ and consider a vector field $X$ which is nonzero and non vertical; hence $V_j \interno (X \interno \d \eta) = 0$ for $j=1,2$ means that $\chi := X \interno \d \eta$ does not contain $\d z$ or $\d w$ factors. However, this is impossible unless $X \interno \d \eta = 0$. Indeed, $\chi \in \Lambda^k (M)$, hence $\chi = F(\xb,z,w) \d x^1 \w ... \w \d x^k$; this cannot be obtained by $\chi = X \interno \beta$ if $X$ is independent of $\pa_z$ and $\pa_w$. 
In other words, we have proven the 

\medskip\noindent
{\bf Lemma 3.} {\it Let $X$ be a non-vertical vector field for the fibration $\pi : M \to B$, where ${\rm dim} (M) = {\rm dim} (B) + 2$. Then $V \interno (X \interno \d \eta) = 0$ for all $V \in \V (\pi)$ implies -- and is thus equivalent to -- $X \interno \d \eta = 0$.}
\medskip

Note that this applies to $W$ provided this is non vertical, i.e. provided the $A^\mu$ identifying $\d \eta$ -- see (10) -- are not all identically vanishing. However, as remarked above, in the case we are discussing $A^1 \not= 0$ at all points, and this condition is hence satisfied.
We have thus proven the following:

\medskip\noindent
{\bf Theorem 2.} {\it The characteristic vector field $W$ for the variational principle defined by $\eta$ satisfies $W \interno \d \eta = 0$.}
\medskip

Recalling that $\N (\d \eta)$ is one dimensional (see also remark 1 for its explicit description) we also have the

\medskip\noindent
{\bf Corollary.} {\it Any vector field $V$ such that $V \interno \d \eta = 0$ satisfies $V = f W$ for some smooth function $f : M \to \R$, with $W$ the characteristic vector field for the variational principle defined by $\eta$.}
\medskip

It follows from lemma 3 that we can redefine proper maximal degree variational principles and their characteristic vector fields of a maximal degree variational principle in a coordinate-independent manner as follows.

\medskip\noindent
{\bf Definition 1.} {\it The maximal degree variational principle defined by $\eta$ on the fiber bundle $\pi : M \to B$ is {\rm proper} if there are vector fields $V_1 , V_2 \in \V (\pi)$ such that $[V_1 \interno (V_2 \interno \d \eta)] \not= 0$.}

\medskip\noindent
{\bf Definition 2.} {\it A vector field $W$ on $M$ satisfying $W \interno \d \eta = 0$ is a {\rm characteristic vector field} for the maximal degree proper variational principle on $\pi : M \to B$ defined by $\eta$.} 
\medskip

Using lemma 3, the discussion of the previous section and that of section 1, when compared, show that the vector field $Z$ defined in (4) is a characteristic vector field for the variational principle defined by $\vth$, see (5); note that imposing $W \interno \d t = 1$ selects among the characteristic vector fields the one with the proper normalization, i.e. imposes $W \equiv Z$.

This provides therefore a unique characterization of the Liouville vector field $X$ in terms of the variational principle on the bundle $\pi : M \to B$ defined by the form $\vth$. We summarize our result in the following

\medskip\noindent
{\bf Theorem 3.} {\it Let $X$ be a vector field on the phase space manifold $P$, globally Liouville w.r.t. the volume form $\Om = \d \s$ and satisfying (1). Then the vector field $Z := \pa_t + X$ on the  enlarged phase space $M = \R \times P$ is a characteristic vector field for the variational principle on $\pi : M \to B$ defined by the form $\vth$ in (3), and is uniquely selected by the normalization condition $Z \interno \d t = 1$.}
\medskip

Obviously, our discussion could as well be read in the opposite direction: consider $M = \R \times P$ where $P$ is $N$-dimensional and orientable; then, given any form $\ga \in \La^{N-2} (P)$, our construction provides a unique vector field $X$ on $P$, which is obviously guaranteed to be Liouville w.r.t. the volume form $\Om := \d \s$. 

This is similar to the situation met in Hamilton dynamics on a symplectic manifold $(S,\om)$, where the relation between the symplectic form $\om \in \La^2 (S)$, the hamiltonian function $H \in \Lambda^0 (S)$ and the hamiltonian vector field $X$ is given by $X \interno \om = \d H$. On the one hand, any given $H$ generates through this relation a unique vector field $X$. On the other hand, one can assign the vector field $X$ and look for functions $H$ (obviously depending on $X$ itself) generating such a vector field; as well known this also provides a variational characterization, via Hamilton's principle, for the vector field $X$. 

\Remark{4.} Finally, we mention that here we have dealt with vector fields which are globally Liouville, see sect.1.1; however, similar results hold in the case of locally Liouville vector fields \cite{Mar}, with obvious modifications: e.g. the forms $\ga$ and $\s$ are not necessarily globally defined, and one has to work chart by chart. Similarly, to avoid chartwise discussion we dealt with the case of globally exact volume form, $\Om = \d \s$, but this restriction is inessential, as discussed at the end of sect.1.1. $\odot$

\section{Examples -- I}

In this section we want to consider three general classes of Liouville dynamics, i.e. Hamilton, Nambu, and Hyperhamilton dynamics; specific equations will be considered in next section. 

Note that, in view of our results (see in particular theorem 2) the interesting object is the form $\vth$, while the choice of the fibration $\pi : M \to B$ is inessential and has no intrinsic meaning (see remark 3); we will thus provide, in this and the next section,  just $\vth$ in order to describe how the considered systems are dealt with in our formalism.

\subsection{Hamilton dynamics}

As well known, any hamiltonian vector field is also Liouville. Let us describe how this is identified by a maximal degree variational principle (beside the standard minimal degree variational principle based on the Poincar\'e-Cartan one-form).

Let $P$ be a smooth manifold of dimension $n=2m$, equipped with a symplectic form $\om$; we write $\zeta = (1/(m-1)!) (\om)^{(m-1)}$ (this is obviously an exterior power). Choose a form $\rho \in \Lambda^1 (P)$ such that locally $\om = m \d \rho$.

The smooth function $H : P \to \R$ defines the hamiltonian vector field $X$ by $X \interno \om = \d H$. On the other hand $\Om = (1/m!) (\om)^{m} = (1/m) \om \w \zeta$. Thus 
$$ X \interno \Om \ = \ (X \interno \om ) \w \zeta \ = \ \d H \w \zeta \ , \eqno(16) $$
and $X$ satisfies (1) with $\ga = H \, \zeta$.  

It follows immediately that the corresponding maximal degree variational principle is based on the form $\vth \in \Lambda^{2m-1} (M)$ (recall $M = \R \times P$) given by 
$$ \vth \ = \ \rho \w \zeta \ + \ H \, \zeta \w \d t \ = \ (\rho \, + \, H \, \d t ) \w \zeta \ ; \eqno(17) $$
the Hamilton equations are readily recovered from this. 

\subsection{Nambu dynamics}

Nambu dynamics \cite{Nam} encountered a renewal of interest in recent years, see e.g. \cite{Ma3,Tak}; see \cite{Mor} for a discussion of it in terms of forms and Cartan ideals.
It is well known that Nambu dynamics is also Liouville, and that in general it cannot be described in terms of a standard (i.e. degree one) variational principle.

An intrinsic definition of Nambu vector fields is as follows: consider a smooth $n$-dimensional manifold $P$ with volume form $\Om$. Then the vector field $X$ on $P$ is Nambu if there is a choice of $n-1$ smooth functions  $H_i : P \to \R$ ($i= 2,...,n$) such that 
$$ \d H_2 \w ... \w \d H_n  \ := \ \ \chi \ = \ X \interno \Om \ . \eqno(18) $$

We have immediately that $\chi = \d \ga$ with e.g. 
$$ \ga \ = \ H_2 \, \d H_3 \w ... \w \d H_n \ . \eqno(19) $$
We also write e.g. $\s = x^1 \d x^2 \w ... \w x^n$, which yields $\d \s = \Om$. (For both $\ga$ and $\s$ one could use a different permutation of indices).

With these, $\th$ is readily recovered, see (3), and hence -- for any $X$ -- we have determined the maximal degree variational principle defining the Nambu vector field $X$.

\subsection{Hyperhamiltonian vector fields}

Another special class of Liouville vector fields is provided by {\it hyperhamiltonian} vector fields, generalizing Hamilton dynamics and  studied in \cite{GM,MT}; these are based on hyperkahler (rather than symplectic) structures \cite{Ati}.

In this case, one considers a riemannian manifold $(P,g)$ of dimension $p=4N$, equipped with three independent symplectic structures $\om_\a$ ($\a = 1,2,3$); to a triple of smooth functions $\h^\a : P \to \R$ one associates a triple of vector fields by (no sum on $\a$) $ X_\a  \interno  \om_\a  =  \d  \h^\a $. The {\it hyperhamiltonian vector field} $X$ on $P$ associated to the triple $\{ \h^\a \}$ is the sum of these, $ X := \sum_{\a=1}^3  X_\a $; it is trivial to check that the $X_\a$, and therefore $X$, are uniquely defined. Each $X_\a$ is obviously Liouville, and so is $X$.
 
On the $(p+1)$ dimensional manifold $M = \R \times P$ (denote by $t$ the coordinate on $\R$) the time evolution under $X$ is described by the vector field $Z = \pa_t + X$. 

Let $\rho_\a$ be one-forms (non unique, and possibly defined only  locally) satisfying $\d \rho_\a = \om_\a$, and $\z_\a$ the $(2N-1)$-th exterior power of $\om_\a$. Define (with $s=\pm 1$ taking care of orientation matters \cite{GM})
$$ \vth \ = \ \sum_{\a=1}^3 \ \rho_\a \w \z_\a \ + \ (6 N s) \, \sum_{\a=1}^3 \ \h^\a \ \z_\a \w \d t \ . \eqno(20) $$

It is immediate to check that $\d \vth$ is nonsingular, and that 
$ Z \interno \d \vth = 0$, $Z \interno \d t = 1$.
It follows from our general discussion that the vector field $Z$ is also obtained by a maximal degree variational principles based on the form $\vth$.

\section{Examples -- II}

In this section we consider some concrete and simple -- but physically relevant -- examples: the Euler top, the ABC flow, and the motion of a charged particle and of a spin in a magnetic field. 

\subsection{The rigid body}

Let us consider the Euler equations for rotations of a free rigid body around its center of mass in three-dimensional space with orthonormal basis vectors $({\bf e}_1 , {\bf e}_2 , {\bf e}_3 )$; we work in the space $R^3$ of angular velocities, so that $x$ represents the angular velocity and $x^i$ is the component of the angular velocity in the direction ${\bf e}_i$. The evolution vector field is now given by $ X = f^i \pa_i $ where
$$ \begin{array}{c}
f^1 := \mu_1 x^2 x^3 \ ; \  f^2 := \mu_2 x^3 x^1 \ ; \ f^3 := \mu_3 x^1 x^2 \\ ~ \\
\mu_1 = (I_2 - I_3)/I_1 \ ; \ \mu_2 = (I_3 - I_1)/I_2 \ ; \ \mu_3 = (I_1 - I_2)/I_3 \ . \end{array} \eqno(21) $$
The volume form is $\Om = \d x^1 \w \d x^2 \w \d x^3$, hence 
$ X \interno \Om = f^1 \d x^2 \w \d x^3 - f^2 \d x^1 \w \d x^3 + f^3 \d x^1 \w \d x^2$.

Writing the one-form $\ga$ as $ \ga = A_\mu \d x^\mu$, equation (1) reads $ f^i = \epsilon_{ijk} \pa_j A_k$ 
or, in vector notation, ${\bf f} = {\rm rot} ({\bf A})$.
The solution to this is, up to an exact form $\gamma' = \d g$ ($g$ a scalar function),
$$ 
A_1 = {1 \over 2} \mu_2 x^1 (x^3)^2 \ , \ 
A_2 = {1 \over 2} \mu_3 x^2 (x^1)^2 \ , \ 
A_3 = {1 \over 2} \mu_1 x^3 (x^2)^2 \ . \eqno(22) $$
Therefore, see (3), the $\th$ appearing in the variational principle for the Euler rigid body equations is (with the $A_i$'s given above), 
$$ \vth \ = \ x^1 \, \d x^2 \w \d x^3 \ - \ \sum_{i=1}^3 A_i \, (\d x^i \w \d t) \ . \eqno(23) $$

\subsection{The ABC flow}

The so called ABC flow has been introduced by Henon and is of interest in fluid dynamics; it lives in $R^3$ is characterized by the fact that, with $X = f^i \pa_i$, the vectors ${\bf f}$ and ${\rm rot} ({\bf f} )$ are collinear (see e.g. \cite{Arn}). 

We have explicitely
$$ \begin{array}{ll}
f^1 = A \sin (x^1) + C \cos (x^2) \ , & \ f^2 = B \sin (x^1) + A \cos (x^2) \\
f^3 = C \sin (x^2) + B \cos (x^1) \ . & ~ \end{array} \eqno(24) $$ 
The volume form is again $\Omega = \d x^1 \w \d x^2 \w \d x^3$, hence 
$$ \begin{array}{rl}
X \interno \Omega \ =& \ [ C \sin (x^2 ) + B \cos (x^1) ] \, \d x^1 \w \d x^2 \\ &  - \ [B \sin (x^1) + A \cos (x^3) ] \, \d x^1 \w \d x^3 \\
& \ + [ A \sin (x^3) + C \cos (x^2) ] \, \d x^2 \w \d x^3 \ .  \end{array} \eqno(25) $$
The solution to $X \interno \Om = \d \ga$ is, up to exact forms $\ga'$, given by 
$$ \begin{array}{rl}
\ga \ =& \ [A \sin (x^3) + C \cos (x^2) ] \d x^1 \, + \, [ B \sin (x^1) + A \cos (x^3) ]  \d x^2 \\
& + \, [C \sin (x^2) + B \cos (x^1) ] \d x^3 \ . \end{array} \eqno(26) $$
The $\vth$ defining our variational principle for the ABC flow is therefore, up to an inessential closed form, given by 
$$ \vth \ = \ x^1 \, \d x^2 \w \d x^3 \ - \ \ga \w d t \ . \eqno(27) $$

\subsection{Particle motion in a stationary magnetic field}
\def\xb{{\bf x}}
\def\Bb{{\bf B}}
\def\wb{{\bf w}}
\def\vb{{\bf v}}

We consider now a point particle of mass $m$ and charge $q$ moving in three-dimensional space under the effect of magnetic field $\Bb$ (no electric field). Then the equation of motions are
$ {\ddot \xb} = k (\vb \times \Bb)$, with $k = q/m$ and with $\times$ the cross (vector)  product in $\R^3$.
We can rewrite these as a first order system in $\R^6$ as
$$ \begin{array}{rl}
{\dot \xb} \ =& \ \vb \\
{\dot \vb} \ =& \ k \ (\vb \times \Bb)  \end{array}
\eqno(28) $$
(if $\Bb$ is also constant in space, we can just consider the second equation and work in the three-dimensional space of velocities).
The corresponding vector field, obviously with zero divergence, is thus 
$$ X \ := \ v^i (\pa / \pa x^i) \ + \ w^i (\pa / \pa v^i ) 
\eqno(29) $$
where we have defined, with $\epsilon$ the Levi Civita tensor, 
$ w^i := k \epsilon^i_{~j\ell} v^j B^\ell $. 

We write $\om_i := \d x^i \w \d v^i $ (no sum on $i$), and the volume form $\Om$ in $\R^6$ will be written as $\Om = \om_1 \w \om_2 \w \om_3$.

We have, with this notation, 
$$ \begin{array}{rl}
X \interno \Om \ =& \ (v^1 \d v^1 - w^1 \d x^1) \w \om_2 \w \om_3 + (v^2 \d v^2 - w^2 \d x^2 ) \w \om_3 \w \om_1 + \\ & \ +  (v^3 \d v^3 - w^3 \d x^3 ) \w \om_1 \w \om_2 ] \ . \end{array} \eqno(30) $$

Let us consider a $\ga \in \Lambda^4 (\R^6)$ of the form
$$ \begin{array}{rl}
\ga \ =& \ \ga_1 \ - \ \ga_2 \ ; \\ 
\ga_1 \ :=& \ (1/2) \ \[ (v^1)^2 \, \om_2 \w \om_3 \ + \ (v^2)^2 \, \om_3 \w \om_1 \ + \ (v^3)^2 \, \om_1 \w \om_2 \] \ , \\ 
\ga_2 \ :=& \ 
[F_a (x) v^2 + F_b (x) v^3 ] \, \om_2 \w \om_3 \ + \ 
[G_a (x) v^1 + G_b (x) v^3 ] \, \om_3 \w \om_1 \\
 & \ + \ [H_a (x) v^1 + H_b (x) v^2 ] \, \om_1 \w \om_2 \ . \end{array} \eqno(31) $$

With these choices, we obviously have that 
$$ \begin{array}{rl} 
\d \ga_1 \ =& \ v^i \left[ (\pa / \pa x^i ) \interno \Om \right]  \ = \\ & = \ v^1 (\d v^1 \w \om_2 \w \om_3) + v^2 (\d v^2 \w \om_3 \w \om_1) + v^3 (\d v^3 \w \om_1 \w \om_2) \end{array} \eqno(32) $$ 
and thus, recalling that $\Bb$ does not depend on $\vb$, $X \interno \Om = \d \ga$ if and only if 
$$ \begin{array}{ccc}
(\pa F_a / \pa x^1 ) = k B^3 &  (\pa F_b / \pa x^1) = - k B^2 &
(\pa G_a / \pa x^2 ) = - k B^3 \\ 
(\pa G_b / \pa x^2) = k B^1 & (\pa H_a / \pa x^3 ) = k B^2 & 
(\pa H_b / \pa x^3) = - k B^1 \end{array} \eqno(33) $$
Needless to say, we can always find functions $F_a , ... , H_b$ satisfying these, just by integrating $B^i$ in different variables $x^j$ ($i,j=1,2,3$).

As for $\s$, we can e.g. choose 
$$ \s \ = \ (1/3) \ ( x^1 \d v^1 \w \om_2 \w \om_3 + x^2 \d v^2 \w \om_3 \w \om_1 + x^3 \d v^3 \w \om_1 \w \om_2 ) \ ; \eqno(34)$$ the form $\vth$ corresponding to $X$ given in (29) is then immediately obtained by (3) using this and (31).

\subsection{Spin motion in a magnetic field.}

As a last example, we consider a particle with spin 1/2 in a spatially constant magnetic field, and the evolution of its spin wave function as described by the Pauli equation. Disregarding evolution of the wave function associated to the spatial coordinates, this reads
$$ {d \Psi \over d t} \ = \ i \, \kappa \ ( {\bf B \cdot S} ) \, \Psi \ . \eqno(35) $$
Here $\kappa = 4 \pi \mu / h$ is a dimensional constant, $\Psi$ is a two-components spinor, $ \Psi = (\psi_+ , \psi_- )$ with $\psi_\pm (t) \in \C$ and $| \Psi |^2 = 1$, the real vector ${\bf B}$ is the magnetic field, with components ${\bf B} (t) = (B_x , B_y , B_z)$, and ${\bf S}$ is the vector spin operator with components the Pauli $\s$ matrices, so that the linear operator $\Mb := {\bf B \cdot S}$ appearing in (35) is given by
$$ \Mb \ = \ \pmatrix{B_z & B_x - i B_y \cr B_x + i B_y & - B_z \cr} \ . \eqno(36) $$

The equation (35) can be rewritten as an equation in $\R^4$ rather than in $\C^2$. In order to do so, we rewrite $\psi_\pm$ separating their real and imaginary part as $\psi_\pm = \chi_\pm + i \z_\pm$.
With the isomorphism $\C^1 \simeq \R^2$ given by $1 \simeq (1,0)$, $i \simeq (0,1)$, the operator of multiplication by $i$ is represented in $\R^2$ by the real antisymmetric matrix with $J_{12} = -1 = - J_{21}$, and thus
$$  i \Mb \approx \pmatrix{B_z J & B_y I + B_x J
\cr - B_y I + B_x J & - B_z J \cr} \ . \eqno(37) $$

Finally, we have obtained that the $\R^4$ representation of equation (31) is given by
$$ {d \xi \over dt} \ = \ \kappa \, A \, \xi \ , \eqno(38) $$
where
$$ \xi \, = \, \pmatrix{\chi_+ \cr \z_+ \cr \chi_- \cr \z_- \cr} \ \ , \ \
A \ = \ \pmatrix{ 0 & - B_z & B_y & - B_x \cr B_z & 0 & B_x & B_y \cr
- B_y & - B_x & 0 & B_z \cr B_x & - B_y & - B_z & 0 \cr} \ \ .
\eqno(39) $$

It is easy to check by explicit computation that this equation is hyperhamiltonian (with orientation corresponding to $s=-1$), with the symplectic structures given by 
$$ \begin{array}{ll}
\om_1 \ = \ \d x^1 \w \d x^3 + \d x^2 \w \d x^4 \ , & 
\om_2 \ = \ \d x^4 \w \d x^1 + \d x^2 \w \d x^3 \ , \\ 
\om_3 \ = \ \d x^2 \w \d x^1 + \d x^3 \w \d x^4 \ ; \end{array} 
\eqno(40) $$
and the hamiltonians given by
$$ \begin{array}{ll}
\h^1 (\xi,t) = (1/2) B_y (t) \| \xi \|^2 \ , & 
\h^2 (\xi,t) = (1/2) B_x (t) \| \xi \|^2 \ , \\ 
\h^3 (\xi,t) = (1/2) B_z (t) \| \xi \|^2 \ . \end{array} \eqno(41) $$
We can choose e.g. 
$$ \rho_1 = x^1 \d x^3 + x^2 \d x^4 \ , \ \rho_2 = x^4 \d x^1 + x^2 \d x^3 \ , \ \rho_3 = x^2 \d x^1 + x^3 \d x^4 \ ; \eqno(42) $$
with these and (40),(41) above, the expression for $\vth$ is immediately read from (20), with $s=-1$. 

\vfill\eject

\section*{Appendix. \\ Hodge duality and the variational principle}

In this appendix we show how the forms entering in our discussion is related to the vector field $X$ via Hodge duality.

Consider a smooth $m$-dimensional orientable riemannian manifold $(M,g_0)$ with local coordinates $\xi^i$ ($i=1,...,m$). In a local chart, vector fields and one-forms admit the bases made of $\pa_i := \pa / \pa \xi^i$ and $\d \xi^i$; these are dual to each other and we write $\pa_i = \widetilde{\d \xi^i}$ and $\d \xi^i = \widetilde{\pa_i}$ to denote this duality. We denote the Levi Civita (completely antisymmetric) covariant tensor by $\epsilon_{\mu_1 ... \mu_m}$, with value $\pm 1$ according to the parity of the permutation $(\mu_1 , ... \mu_m)$.

The Hodge star is the linear map $* : \Lambda^r (M) \to \Lambda^{m-r} (M)$ defined by 
$$ * \ (\d \xi^{\mu_1} \w ... \w \d \xi^{\mu_r} ) \ := \ { \sqrt{|g_0|} \over (m-r)!} \ {\epsilon^{\mu_1 ... \mu_r}}_{\nu_{r+1} ... \nu_{m} } \ \d \xi^{\nu_{r+1}} \w ... \w \d \xi^{\nu_m} \ . $$

Let, as in the main body of the paper, $P$ be a smooth orientable riemannian manifold of dimension $N$ with metric $g$ and volume form $\Om$, on which a vector field $X$ satisfying (1) for some $\ga \in \La^{N-2}(P)$ is defined. We consider the manifold $M = \R \times P$, with metric $g_0 = \delta \otimes g$ (so that $|g_0| = |g|$) and coordinates $(t,x^1,...,x^N)$.

Let us now consider the form $\vth$ defining the variational principle for $X = f^i \pa_i$, see theorem 3 above, and given by (3). We have $ \vth = \s + \d t \w \ga$ and thus $\d \vth  = \Om + \d t \w \d \gamma$. Recalling (1), we have $\d \vth = \Om + f^i [\d t \w (\pa_i \interno \Om)]$; the Hodge dual of this is
$$ * (\d \vth) \ = \ *(\Om) + f^i *[(\pa_i \interno \Om) \w \d t] \ = \ \sqrt{|g^{-1}|} \[ \d t + \sum_{j=1}^N f^j \d x^j  \] \ . $$ 
If now we consider the vector field which is dual of this one-form in the sense of the duality between forms and vector fields, we have
$$ \widetilde{* (\d \vth)} \ = \ \sqrt{|g^{-1}|} \ \[ \pa_t + X \] \ = \ \sqrt{|g^{-1}|} \ Z \ . $$
By reversing the argument, or by acting first with the tilde and then the star operators on this equation, we have that:

\medskip\noindent
{\bf Lemma.} {\it The form $\vth \in \La^{N-1}(M)$ defining the variational principle associated to the Liouville vector field $X$ satisfies $ \d \vth \ = \ \sqrt{|g^{-1}|} \ *(\widetilde{Z})$.}
\medskip

Note that this condition completely determines the variational principle: indeed it identifies $\vth$ up to a closed form, which has no role in the variation of $I(\phi) = \int_D \phi^* (\vth )$.

\vfill\eject

\end{document}